\documentclass{PoS}

\title{What's up with IR gluon and ghost propagators \\
       in Landau gauge? A puzzling answer from \\ huge lattices}

\ShortTitle{IR propagators A puzzling answer from huge lattices}

\author{\speaker{Attilio Cucchieri} and Tereza Mendes \\
        Instituto de F\'\i sica de S\~ao Carlos, Universidade de S\~ao Paulo, \\
        Caixa Postal 369, 13560-970 S\~ao Carlos, SP, Brazil \\
        E-mail: \email{attilio@ifsc.usp.br}, \email{mendes@ifsc.usp.br}}

\abstract{Several analytic approaches predict for $SU(N_c)$ Yang-Mills theories
in Landau gauge an enhanced ghost propagator $G(p^2)$ and a suppressed gluon
propagator $D(p^2)$ at small momenta. This prediction applies to two, three and four
space-time dimensions. Moreover, the gluon propagator is predicted to be null at
$p = 0$. Numerical studies by several groups indeed support an enhanced ghost
propagator when compared to the tree-level behavior $1/p^2$ and a finite infrared
gluon propagator. However, the agreement between analytic and numerical studies
is only at the qualitative level in three and in four dimensions. In particular,
the infrared exponent of the ghost propagator seems to be smaller than the one
predicted analytically and the gluon propagator seems to display a (finite)
nonzero value at zero momentum. It has been argued that this discrepancy might
go away once simulations are done on much larger lattice sizes than the ones
used up to now. Here we present data in three and four space-time dimensions
using huge lattices in the scaling region, i.e.\ up to $320^3$ at $\beta = 3.0$
and up to $128^4$ at $\beta = 2.2$, corresponding to $V \approx (85 \,\mbox{fm})^3$ and
$V \approx (27 \,\mbox{fm})^4$.
Simulations have been done on the IBM supercomputer at the University of S\~ao
Paulo.}

\FullConference{The XXV International Symposium on Lattice Field Theory\\
		 July 30-4 August 2007\\
		 Regensburg, Germany}

\begin{document}


\section{Introduction}

In the Gribov-Zwanziger confinement scenario for Landau gauge
\cite{Gribov:1977wm,Zwanziger:1991gz} the gluon
propagator $D(p^2)$ is predicted to be infrared (IR) suppressed at small
momenta. In particular, one should have $D(0) = 0$, implying that
reflection positivity is maximally violated. This violation of
reflection positivity may be viewed
as an indication of gluon confinement \cite{Alkofer:2000wg}. At the same
time, the Gribov-Zwanziger \cite{Gribov:1977wm,Zwanziger:1993dh}
and Kugo-Ojima \cite{Kugo:1979gm} confinement scenarios predict (in
Landau gauge) a ghost propagator $G(p^2)$ enhanced in the IR limit.
This represents a long-range effect and could be related to quark confinement
\cite{Gribov:1977wm,Alkofer:2000wg,Zwanziger:1993dh,conf}.

Several analytic studies \cite{von Smekal:1997vx,SD,Lerche:2002ep,Braun:2007bx}
agree with the above scenarios predicting, for small momenta, a gluon propagator
$D(p^2) \propto p^{2(a_D - 1)}$ and a ghost propagator $G(p^2) \propto
1/ p^{2(1 + a_G)}$, with the relation $ a_D = 2 a_G + (4 - d)/2$. Here
$d$ is the space-time dimension. Clearly, if $a_D > 1$ one has $D(0) = 0$.
In the four-dimensional case, one finds \cite{SD,Lerche:2002ep}
$a_G \approx 0.59$ and $a_D = 2 a_G$.
Similar power behaviors have also been obtained for the various vertex
functions of $SU(N_c)$ Yang-Mills theories \cite{Lerche:2002ep,Fischer:2006vf,
Fischer:lat07}.

These results have been confirmed at the quantitative level in the
two-dimensional case, using lattices up to almost $(43 \,\mbox{fm})^2$ \cite{Maas:2007uv}.
In the three-dimensional case \cite{Cucchieri:2003di},
one clearly sees an IR-suppressed gluon propagator. However, using lattice
volumes up to about $(24 \,\mbox{fm})^3$, it was not possible to control the extrapolation
to infinite volume. In particular, the data for the rescaled gluon propagator
at zero momentum $D(0)$ could be fitted [as a function of the inverse lattice
side $1/L = 1/( N a )$] using the Ansatz $d + b / L^c$ both with $d = 0$ and with $d \neq 0$
\cite{Cucchieri:2003di}. Here $a$ is the lattice spacing and $N$ is the
number of lattice points per direction.
Finally, in four dimensions, the gluon propagator
is clearly less divergent than in the tree-level case \cite{gluon,
Cucchieri:1997dx,Ilgenfritz:2006he,Cucchieri:2006xi}.
On the other hand, even using lattices with a lattice side of about $10 \,\mbox{fm}$,
one does not see a gluon propagator decreasing at small momenta \cite{Cucchieri:2006xi}.
One should stress, however, that the Landau gluon propagator clearly
violates reflection positivity, in two, three and four space-time
dimensions \cite{Maas:2007uv,Cucchieri:2004mf,violation}.
For the ghost propagator, the IR exponent $a_G$ obtained using analytic
studies has been confirmed in 2d \cite{Maas:2007uv}, while in the 3d
\cite{Cucchieri:2006tf} and in the 4d \cite{Cucchieri:1997dx,Ilgenfritz:2006he,
ghost,Boucaud:2005gg}
cases the exponent obtained using lattice numerical
simulations is always smaller than the one predicted analytically.
Let us also recall that, in Ref.\ \cite{Cucchieri:2007zm}, it was
shown that gluon and ghost propagator for $SU(2)$ and $SU(3)$ Yang-Mills
theories are in very good agreement from momentum $p \approx 1$ GeV
to about $p \approx 10$ GeV. Similar results have been presented in \cite{AGW-LAT07}.
These findings suggest that the IR behavior of these propagators is
independent of the gauge group $SU(N_c)$, as predicted analytically
\cite{von Smekal:1997vx,SD,Lerche:2002ep}.

Finally, by solving  Dyson-Schwinger equations on a finite four-dimensional
torus \cite{Fischer:lat07,Fischer:2002eq,Fischer:2007pf} one
can show that the gluon propagator (at small momenta) seems to diverge for
volumes up to about $(8 \,\mbox{fm})^4$, develops a plateau for $V \approx (9 \,\mbox{fm})^4$
and is IR suppressed for a lattice side larger than $10 \,\mbox{fm}$. Also, the
extrapolation of these results to infinite volume gives a null gluon propagator
at zero momentum. At the same time, one obtains that, after eliminating (for
each volume) the data corresponding to the first two non-zero momenta, the
ghost propagator shows a power-law behavior and, in the infinite volume-limit,
one obtains the IR exponent $a_G$ predicted by analytic studies
\cite{von Smekal:1997vx,SD,Lerche:2002ep}.

From the above results, it seems necessary to extend present numerical
simulations to very large lattice volumes in order to verify if the
agreement obtained between lattice data and analytic results for the
two-dimensional case \cite{Maas:2007uv} applies to the 3d and 4d cases as well.
Here we present extensive simulations in three and in four dimensions,
for the $SU(2)$ case, using huge lattices. In particular, we considered
lattice sides $N = $ 140, 200, 240 and 320 in 3d at $\beta = 3.0$ and $N = $
48, 56, 64, 80, 96 and 128 in 4d at $\beta = 2.2$. In 3d, the number of configurations
was about 630, 525, 350 and 45, respectively for the four lattice sizes, both
for the gluon and for the ghost propagators. In the 4d case we have considered
for the gluon propagator 168 configurations for $V = 128^4$ and about
250 configurations for the other lattice sizes. For the ghost propagator we have 21
configurations for the largest volume and about 100 in the other cases.
For the inversion of the Faddeev-Popov matrix we used the so-called
{\em point-source} method \cite{Cucchieri:2006tf,Boucaud:2005gg}.
Note that the lattice volumes $320^3$ at $\beta = 3.0$
and $128^4$ at $\beta = 2.2$ correspond, respectively, to $V \approx (85 \,\mbox{fm})^3$
and $V \approx (27 \,\mbox{fm})^4$. (See Refs.\ \cite{Cucchieri:2003di,Bloch:2003sk}
for details about how the physical lattice spacing $a$ has been set in the
two cases.) Also note that all our runs are in the scaling region.

These simulations have been done in the IBM supercomputer at USP. This
machine has 112 blades with 2 dual-core PowerPC 970 2.5GHz CPU's,
a Myrinet network and about 4.5 Tflops peak-performance (occupying
position number 363 in the {\tt TOP500} list of November 2006).


\section{Results}

\begin{figure}
\begin{center}
\includegraphics[width=.6\textwidth]{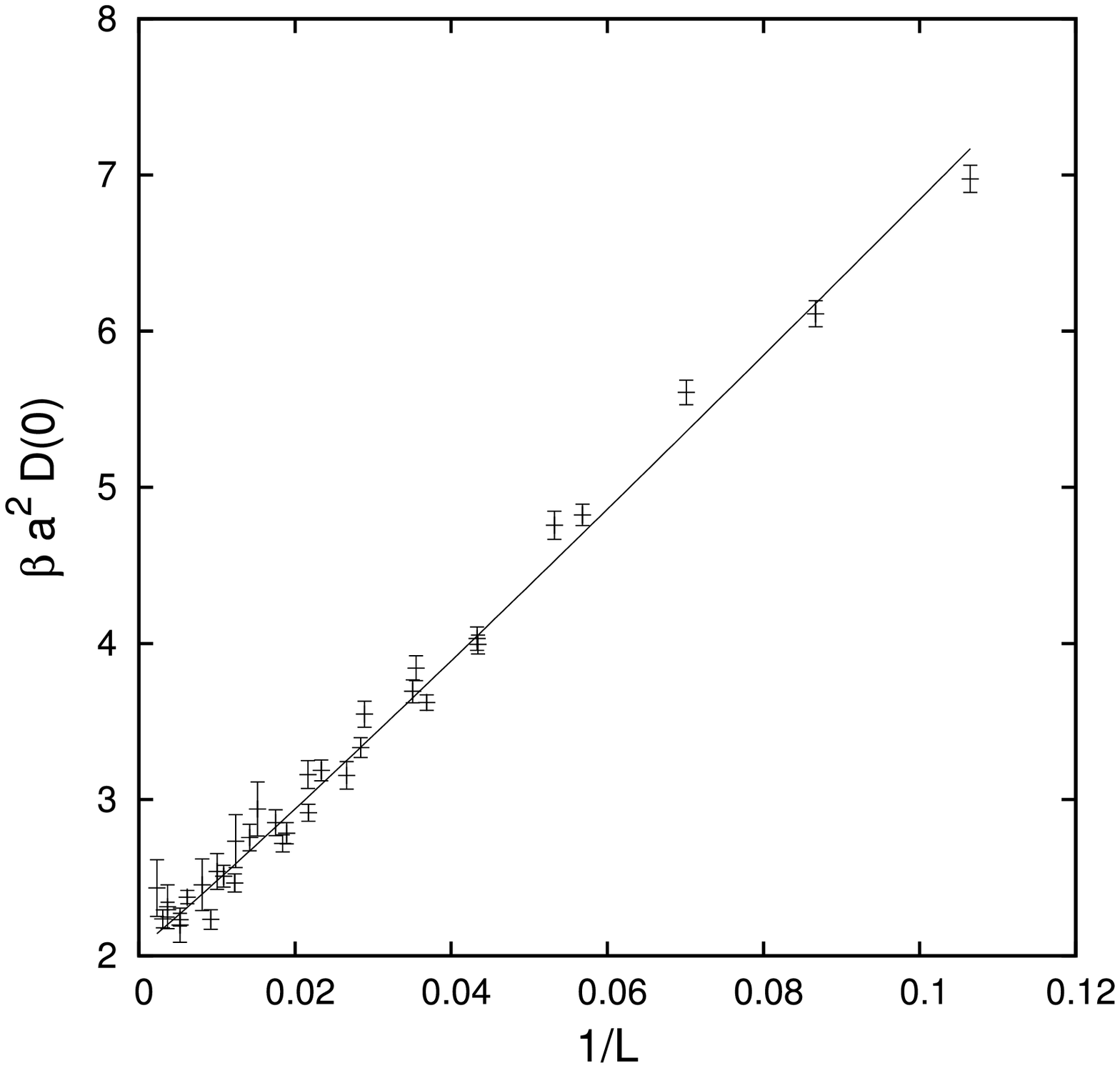}
\end{center}
\caption{Renormalized gluon propagator at zero momentum $\beta a^2 D(0)$
         (in GeV$^{-2}$) as a function of the inverse lattice side $1/L$
         (in GeV) and extrapolation to infinite volume. The fit is given
         by $b + c / L^e$ with $e = 1.04 (5)$ and $b = 2.05(5) \, $ GeV$^{-2}$.
  }
\label{fig:gluon-3d-extrap}
\end{figure}

Considering the new data produced in the 3d case together with
old data from Refs.\ \cite{Cucchieri:2003di,Cucchieri:2004mf},
we tried an extrapolation to infinite volume for the gluon
propagator $D(0)$ as a function of the inverse lattice side $1/L$.
The data for the propagator have been renormalized following Ref.\
\cite{Maas:2007uv}. Results are shown in Fig.\ \ref{fig:gluon-3d-extrap}.
It is clear from the plot that there is no sign of a propagator
going to zero as $L$ goes to infinity. Moreover, the data show
a behavior of the type $D(0) \sim 1/L$ and an infinite-volume extrapolation
given by $D(0) \approx 2$ GeV$^{-2}$. This implies $a_D = 1$.

\begin{figure}
\begin{center}
\includegraphics[width=.45\textwidth]{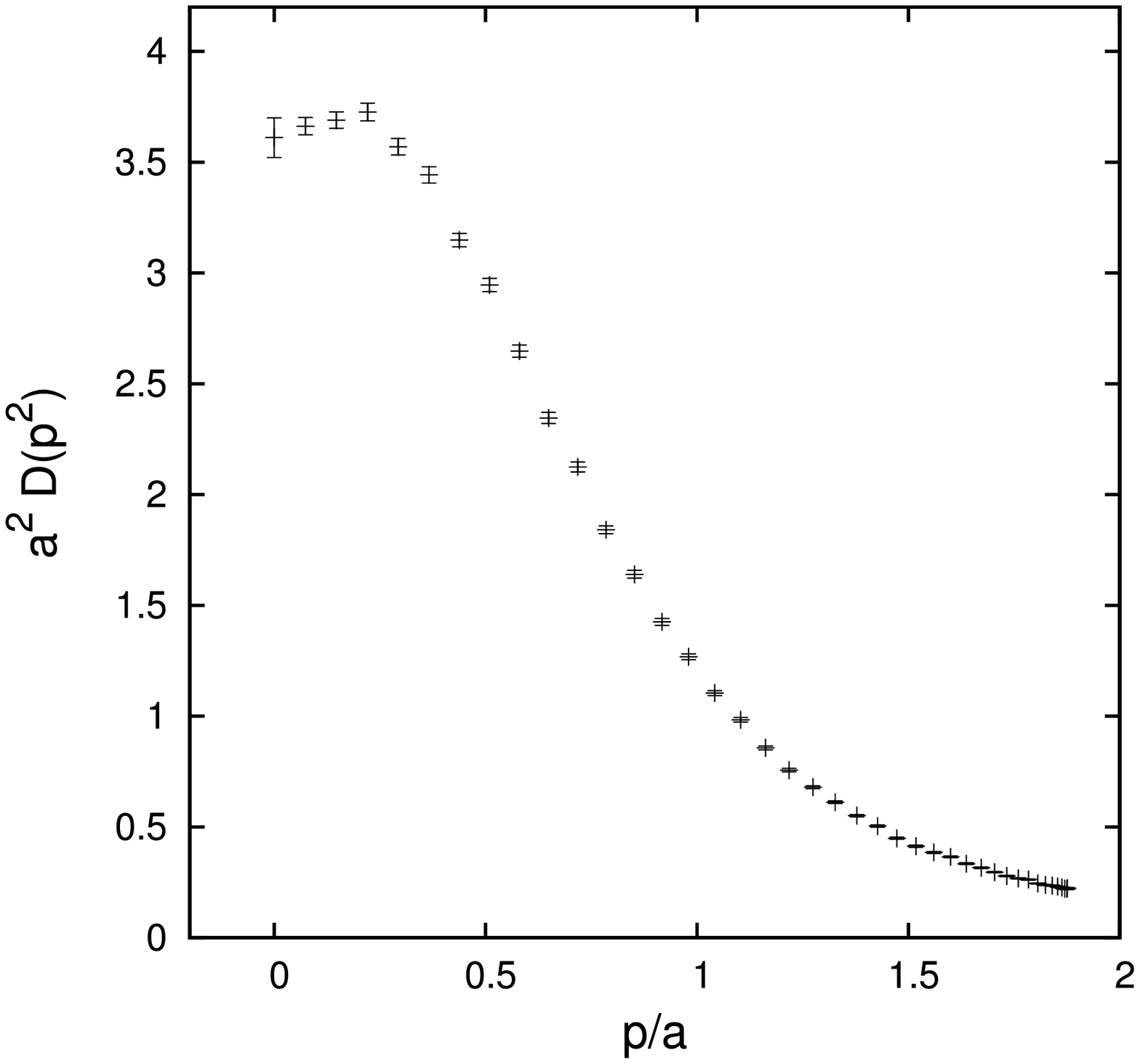}
\hspace{1cm}
\includegraphics[width=.45\textwidth]{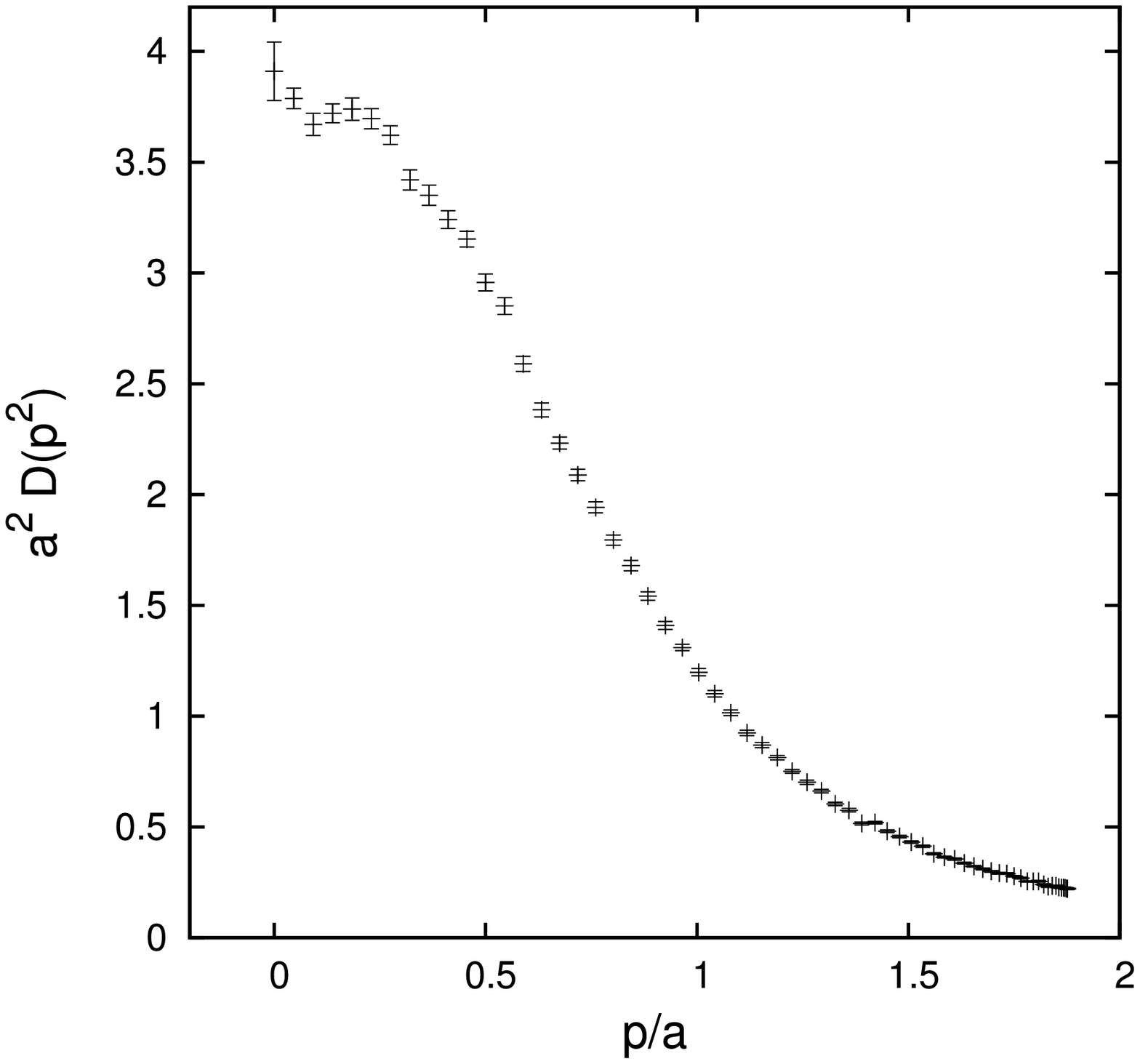}
\end{center}
\caption{Unrenormalized gluon propagator $a^2 D(p^2)$ (in GeV$^{-2}$) as a
         function of the momentum $p/a$ (in GeV) for lattice volumes
         $V = 80^4$ (left) and $V = 128^4$ (right) at $\beta = 2.2$.}
\label{fig:gluon-4d-80-128}
\end{figure}

In the 4d case (see Fig.\ \ref{fig:gluon-4d-80-128}), even considering
very large lattice volumes and relatively large statistics, one cannot
see a clear sign of a gluon propagator $D(p^2)$ decreasing at small
momenta. Similar results have been presented at this conference by other
groups \cite{AGW-LAT07,B-LAT07}. 
Clearly, also in this case the data suggest $a_D = 1$.
On the other hand, violation of reflection
positivity is confirmed and the gluon propagator, considered
as a function of the spatial separation $s$, becomes negative at
$s \approx 1 \,\mbox{fm}$, in agreement with Ref.\ \cite{Alkofer:2003jj}.

We have also tried to estimate the IR exponent $a_G$ for the ghost
propagator (in the 3d and 4d cases) using the Ansatz $G(p) = c/p^{2(1+a_G)}$
and considering, for each lattice volume, either the two smallest nonzero
momenta or the third and fourth smallest nonzero momenta. Results are reported
in Tables \ref{tab:kappa-3d4d}. As one can see, this IR exponent seems
to go to zero as the infinite-volume limit is approached, in agreement with
\cite{Cucchieri:1997dx}. One should however notice that, for $p \approx 500$ MeV, the
exponent $a_G$ is about 0.3, also in agreement with Ref.\
\cite{Cucchieri:1997dx}.

\begin{table}
\begin{center}
\begin{tabular}{ccccccccccc}
 $N^3$ & $a_G$ & \phantom{oi} & $N^4$    & $a_G$ & \phantom{oioioi} &
 $N^3$ & $a_G$ & \phantom{oi} & $N^4$    & $a_G$ \\
\hline
 $140^3$ & 0.073(4) & & $48^4$ & 0.093(7) & &
 $140^3$ & 0.13(2)  & & $48^4$ & 0.19(4)  \\
 $200^3$ & 0.051(3) & & $56^4$ & 0.063(6) & &
 $200^3$ & 0.06(2)  & & $56^4$ & 0.18(4)  \\
 $240^3$ & 0.003(3) & & $64^4$ & 0.049(9) & &
 $240^3$ & 0.10(2)  & & $64^4$ & 0.17(4)  \\
 $320^3$ & -0.021(9) & & $80^4$ & 0.052(5) & &
 $320^3$ &  0.01(5)  & & $80^4$ & 0.15(2)   \\
         &           & & $112^4$ & 0.038(6) & &
         &           & & $112^4$ & 0.10(7)  \\
         &           & & $128^4$ & 0.016(5) & &
         &           & & $128^4$ & 0.06(3) \\
\end{tabular}
\end{center}
\caption{Table for the ghost propagator IR exponent $a_G$, in the 3d and 4d cases,
         obtained using either the two smallest nonzero momenta (left) or
         the third and fourth smallest nonzero momenta (right).}
\label{tab:kappa-3d4d}
\end{table}


\section{Conclusions}

The above results leave us with several open questions. From the
lattice point of view, one should of course investigate if Gribov-copy
effects and/or finite-volume effects could explain our results.  
Let us recall that an improved gauge-fixing method 
\cite{Bogolubsky:2007bw} seems capable of reducing finite-volume
effects for the gluon propagator by enlarging the set of allowed
gauge transformations. At the same time, one needs to reconcile
the above results with the non-renormalizability of the ghost-gluon
vertex \cite{Cucchieri:2004sq} and with the suppression of $D(p^2)$ when
considering simulations in the strong-coupling regime \cite{Cucchieri:1997fy},
in the interpolating gauge (or $\lambda$-gauge) \cite{Cucchieri:2007uj}
and in Coulomb gauge \cite{Cucchieri:2000gu}.
From the analytic point of view, it may seem necessary to
reconsider partially the conventional confinement scenarios
\cite{Gribov:1977wm,Zwanziger:1991gz,Alkofer:2000wg,
Zwanziger:1993dh,Kugo:1979gm,conf,von Smekal:1997vx,SD,Lerche:2002ep,
Braun:2007bx,Fischer:2006vf,Fischer:lat07} discussed in the Introduction.
One should of course
recall that there are different solutions of Dyson-Schwinger equations
for gluons and ghosts in Landau gauge \cite{Aguilar:2004sw,Boucaud:2005ce}.
In particular, the results obtained in Ref.\ \cite{Aguilar:2004sw}
support a finite non-zero gluon propagator and an essentially tree-level
ghost propagator at small momenta. Similar results are obtained
in Ref.\ \cite{Frasca:2007uz}. Phenomenological tests \cite{Natale:2006nv}
also seem to favor $D(0) > 0$.

We believe that a clarification of the present status of the
Kugo-Ojima/Gribov-Zwanziger scenario will probably require new
ideas and new methods, both for analytic and numerical studies,
and that a key point will be a better understanding of the
gauge interpolating between the Landau and the Coulomb gauge
\cite{Fischer:2005qe}.


\section{Acknowledgements}

The authors thank R. Alkofer, C. Fischer, H. Gies,
F.J. Llanes-Estrada, A.A Natale, O. Oliveira, K. Schwenzer and
S. Sorella for discussions
and the Institute of Physics of the University of Graz for
hospitality. We also acknowledge partial support from
FAPESP, CNPq and CCInt-USP. The simulations reported
here have been done on the IBM supercomputer at S\~ao
Paulo University (FAPESP grant \# 04/08928-3).



\end{document}